\documentstyle[11pt,epsf]{article}

\newcommand{\sect}[1]{ \section{#1} \setcounter{equation}{0} }

\newcommand{\partialslash}{\partial \! \! \! /}

\newcommand{\kslash}{k \! \! \! /}

\newcommand{\Dslash}{D \! \! \! \! /}
\newcommand{\xslash}{x \! \! \! /}

\newcommand{\half}{\mbox{\small{$\frac{1}{2}$}}}

\newcommand{\quarter}{\mbox{\small{$\frac{1}{4}$}}}

\newcommand{\threehalves}{\mbox{\small{$\frac{3}{2}$}}}

\newcommand{\MS}{\overline{\mbox{MS}}}
\newcommand{\MMS}{\overline{\mbox{\small{MS}}}}

\newcommand{\NN}{{\cal{N}}}
\newcommand{\eeq}{\end{equation}}
\newcommand{\beq}{\begin{equation}}
\newcommand{\bear}{\begin{eqnarray*}}
\newcommand{\enar}{\end{eqnarray*}}

\begin{document}
\title{On the {$\NN$} $=$ $2$ supersymmetric $CP(N)$ $\sigma$ model and \\
Chern Simons terms.}
\author{Massimiliano Ciuchini, \\ Dipartimento di Fisica dell'Universit\`a \&
INFN, \\ Piazza Torricelli 2, \\ I-56126 Pisa,\\ Italy. \\ \\ and \\ \\
J.A. Gracey, \\ Department of Mathematical Sciences, \\ University of Durham,
\\ South Road, \\ Durham, \\ DH1 3LE, \\ United Kingdom.\thanks{Address after
1st May, 1995: DAMTP, University of Liverpool, P.O. Box 147, Liverpool, L69
3BX, United Kingdom.}}
\date{}
\maketitle
\vspace{2cm}
\noindent
{\bf Abstract.} We use the large $N$ self consistency method to compute the
critical exponents of the fields and coupling of the supersymmetric $CP(N)$
$\sigma$ model at leading order in $1/N$ in various dimensions. We verify that
the correction to the critical $\beta$-function slope vanishes at $O(1/N)$
which is consistent with supersymmetry. The three dimensional model is
investigated explicitly when a Chern Simons term is included
supersymmetrically. We determine the  modification that this has on the gauge
independent quantity $\beta^\prime(g_c)$ as a function of the Chern Simons
coupling, $\vartheta$. For an $\NN$ $=$ $2$ supersymmetric Chern Simons term
the exponent is independent of $\vartheta$ at $O(1/N)$, whilst it is invariant
under $\vartheta$ $\rightarrow$ $1/\vartheta$ when an $\NN$ $=$ $1$
supersymmetric Chern Simons term is included.

\vspace{-20cm}
\hspace{11.5cm}
{\bf {IFUP-TH 31/95}}

\vspace{0.2cm}
\hspace{12.9cm}
{\bf {LTH-353}}
\newpage
\sect{Introduction.}
The supersymmetric $\sigma$ model \cite{1,2,3} defined on the symmetric
K\"{a}hler manifold $CP(N)$ possesses many interesting classical and quantum
properties aside from being a $U(1)$ gauge theory invariant under $\NN$ $=$ $2$
supersymmetry. It has been widely studied from varying points of view. In two
dimensions the classical model possesses instantons, \cite{4}, as well as being
an integrable field theory, which implies the existence of an exact $S$-matrix,
\cite{5}. The absence of an anomaly in the conservation of the quantum
non-local charges of the model, \cite{6}, implies this $S$-matrix is valid in
the quantum theory. More interesting properties emerge upon considering the
quantum corrections to the model. For instance, the model is an asymptotically
free theory in two dimensions. (See, for example, \cite{7}.) Moreover, the
$\beta$-function has been determined exactly to {\em all} orders by instanton
methods in \cite{8}. In fact there are no contributions to $\beta(g)$ beyond
the first order, where $g$ is the coupling. Subsequent explicit perturbative
calculations on arbitrary target manifold to high orders have always been
consistent with this result, [7,9-12]. Indeed it is the presence of the
extended supersymmetry which is responsible for this feature in much the same
way that certain four dimensional super Yang-Mills theories have simple
$\beta$-functions. (See, for example, the review in \cite{13}.) Another feature
is the generation of masses through dynamical symmetry breaking. Moreover, this
mass is non-perturbative in nature ie non-analytic in the coupling and can only
be accessed via non-perturbative techniques such as the large $N$ expansion,
[2,3,14-16]. Interestingly an exact expression for this mass has recently been
deduced in \cite{17} from the exact S-matrix and expressed in terms of
$\Lambda_{\MMS}$. Indeed many authors have analysed the model in large $N$ to
reveal this rich structure either in a manifestly supersymmetric formulation
involving superfields, \cite{14,15,16}, or for the component lagrangian,
\cite{2,3}. However, in the latter approach the choice of Wess Zumino gauge in
defining superfield components leads to a lagrangian where supersymmetry is
broken, though supersymmetric (and gauge independent) results can still be
deduced. Several interesting features emerge in the large $N$ approach. One
which is specific to $\NN$ $=$ $2$ theories is the generation of central
charges in the supersymmetry algebra which are topological in nature,
\cite{2,18}. This was noticed in other models in either two or four dimensions
in \cite{19,20}. The same feature occurs in the three dimensional
supersymmetric $CP(N)$ model, \cite{21}. Their presence is easy to deduce given
that the irreducible representations, (irreps), of the classical model are the
massless ones of dimension $2$. Whilst in the quantum theory to preserve this
dimension of irrep when the fields are then massive, one needs a
supersymmetry algebra with central charges, whose values ordinarily are
bounded by the mass of the irrep. Further, the irrep dimension can only be
preserved precisely when this bound is exactly satisfied, ie saturated. For
other models with topological objects such as solitons, the analogous
condition would correspond to the Bogomolny bound, \cite{19}. Another
important feature which occurs in precisely two dimensions in the context of
mass generation is role the axial anomaly plays in ensuring the fields of one
supermultiplet all have the same mass to be consistent with supersymmetry,
\cite{2}.

One issue of the quantum theory is that of renormalizability. Perturbatively in
two dimensions the model is renormalizable but this is not preserved beyond two
dimensions as can be seen by simple power counting. Instead within the
conventional large $N$ expansion approach one can probe the two and three
dimensional models as then this expansion is renormalizable, \cite{22}. Indeed
both the bosonic and supersymmetric models have been examined in three
dimensions using this approach, \cite{22}, as well as the related $\sigma$
model on the Grassmann manifold $SU(M+N)/[SU(M)\times SU(N)]$, \cite{23,24}.
This involves computing the effective action in the saddle point approximation
when $N$ is large. However, this is not the only way in which to examine models
in the large $N$ expansion. An elegant formulation of analysing models well
beyond the leading saddle point approximation has been available following the
series of papers of Vasil'ev et al which determined information on the $O(N)$
bosonic $\sigma$ model at $O(1/N^3)$, \cite{25,26}. In that approach one can
probe well beyond the leading order by exploiting the scaling or conformal
symmetry available at the $d$-dimensional gaussian fixed point. There the
absence of the mass, which prevents one from proceeding beyond the leading
order saddle point, means that critical exponents can be deduced in arbitrary
dimensions in a power series in $1/N$. These exponents which are directly
related through the critical renormalization group equation to, for example,
the $\beta$-function characterize the quantum properties of the theory. This
relationship allows one to deduce the coefficients appearing in the
perturbative series of $\beta(g)$, at each level in $1/N$ but all orders in the
coupling. Moreover, an advantage of the procedure is that being arbitrary
dimensional one can deduce information on the two and three dimensional models
simultaneously. More recently, the methods of \cite{25} have been successfully
applied to models with $\NN\,=\,~1$ supersymmetry to $O(1/N^2)$, \cite{27},
$O(N)$ four fermi theories to $O(1/N^3)$, \cite{28,29,30}, and the minimal
extension of the $CP(N)$ $\sigma$ model to include fermions, \cite{31}, as well
as various four dimensional gauge theories, \cite{32}. Given the success of the
method in various models the purpose of this paper is to apply the technique to
the supersymmetric $CP(N)$ $\sigma$ model. The aim is to compute the critical
exponents of all the fields, coupling and a class of composite operators. One
issue to be dealt with relates to the $\beta$-function. As the model is a
$U(1)$ gauge theory only those exponents which are gauge independent, like
$\beta^\prime(g_c)$, are physically relevant, where $g_c$ is the critical
coupling constant defined to be the non-trivial zero of the $d$-dimensional
$\beta$-function. In the $\NN$ $=$ $1$ supersymmetric $\sigma$ model on the
$N$-sphere, it turned out that $\beta^\prime(g_c)$ vanished at $O(1/N)$ in
arbitrary dimensions, \cite{12,33}. This was a consequence of the (unbroken)
supersymmetry of the model. Indeed this is consistent with the observation in
perturbation theory that two dimensional $\sigma$ models with $\NN$ $=$ $1$
supersymmetry have no two or three loop terms in their $\beta$-function on any
target manifold, \cite{7}. This vanishing does not persist beyond three loops,
except when more symmetries are present. For instance, it is well known that a
four loop counterterm exists, which vanishes on symmetric K\"{a}hler spaces,
\cite{9}, such as $CP(N)$, consistent with \cite{8}. Therefore, one aim in our
analysis is to examine the corresponding exponent, $\beta^\prime(g_c)$, to
ascertain whether its $O(1/N)$ correction vanishes as in the $O(N)$ case and to
be another check on the instanton result of \cite{8}. Indeed our analysis
builds on the approach of \cite{3,4,5} for the bosonic $CP(N)$ $\sigma$ model.
There is one crucial difference in this paper, however, compared to previous
large $N$ critical exponent analyses in arbitrary dimensions. This relates to
the presence of supersymmetry which, as is well known, is broken when one is
away from integer dimensions. Therefore whilst we will be able to derive
arbitrary dimensional expressions, these will not be meaningful until specific
dimensions are examined when one ought to observe the restoration of
supersymmetry. This will then manifest itself in exponents within the same
supermultiplet becoming equal.

Despite this apparent complication another problem which will be investigated
is the effect that the inclusion of a Chern Simons term in the strictly three
dimensional model has on the results, \cite{36}. For instance, one question to
answer is if the correction to $\beta^\prime(g_c)$ vanishes in three dimensions
without such a topological term, does its inclusion alter this result or is the
critical slope correction non-vanishing. This problem is interesting for
several reasons. First, there has been recent activity in probing models with
Chern Simons terms in the large $N$ expansion. For example, the bosonic $CP(N)$
$\sigma$ model was examined independently in \cite{37} using conventional
methods. There a non-zero correction to the critical $\beta$-function slope as
well as other exponents were determined as a function of $\vartheta$, where
$\vartheta$ is the (dimensionless) coupling of the Chern Simons term which does
not get renormalized, \cite{38}. Primarily the motivation for that work was
then in relation to models with high $T_c$ superconductivity. However, Ferretti
and Rajeev proceeded to examine the current algebra of the model together with
its Grassmannian companion in \cite{39}. Their aim was to gain an insight into
the fixed point structure and to then deduce the theory to which it was
equivalent. Our calculation is an analogous step in that direction for the
supersymmetric case.

The paper is organised as follows. Section two contains useful background on
the model as well as introducing our notation. The basic formalism is
presented, in section three, to deduce the exponents of all the fields as well
as $\beta^\prime(g_c)$. The extension of this work to $\vartheta$ $\neq$ $0$ is
performed in section 4, whilst we draw our conclusions and indicate future work
in section 5.

\sect{Background.}
There are various ways in which the supersymmetric $CP(N)$ $\sigma$ may be
formulated. For example, the action may be written in a manifestly $\NN$ $=$
$2$ supersymmetric fashion in terms of a vector and chiral superfield $V$ and
$R$ respectively, as \cite{3},
\begin{equation}
S ~=~ \int \, d^2x \, d^2\theta \, d^2 \bar{\theta} \, [ R \bar{R} e^V ~-~ V ]
\end{equation}
Here $\theta$ and $\bar{\theta}$ are the complex Grassmann coordinates of
superspace, $V$ $=$ $V(x,\theta,\bar{\theta})$ is a real superfield and $R$ $=$
$R(x_\mu - \half \bar{\theta} \gamma_\mu \theta, \theta)$ satisfies the
chirality condition $D_\alpha R$ $=$ $0$ where $D_\alpha$ is a supercovariant
derivative acting on the full superspace. The action (2.1) is also invariant
under the infinitesimal $U(1)$ gauge transformations
\[
\delta R ~=~ i \Lambda R ~~~,~~~ \delta V ~=~ - \, i ( \Delta - \bar{\Delta} )
\]
where $\Lambda$ is a chiral gauge function. Alternatively the action can be
written in terms of $\NN$ $=$ $1$ superfields by supersymmetrizing the bosonic
model in much the same way as the $O(N)$ supersymmetric $\sigma$ model is
constructed from the bosonic model, \cite{3,15,16,18,40}. Therefore
introducing $\NN$ $=$ $1$ superfields $\Phi$, $\bar{\Phi}$, $\Lambda$ and
$A_\alpha$, where the first two are complex and the remaining ones real with
the latter a spinor, then the action is, \cite{3,18},
\begin{equation}
S ~=~ \int \, d^2 x \, d^2\theta \, [ (D - iA)\bar{\Phi}^i (D+iA)\Phi^i
{}~+~ \Lambda (\bar{\Phi}^i\Phi^i - 1/g) ]
\end{equation}
where $g$ is the coupling constant, $1$ $\leq$ $i$ $\leq$ $N$ and $D_\alpha$ is
the $\NN$ $=$ $1$ supercovariant derivative,
\begin{equation}
D_\alpha ~=~ \frac{\partial ~}{\partial \theta^\alpha} ~-~
i (\gamma^\mu \theta)_\alpha \frac{\partial ~}{\partial x^\mu}
\end{equation}
Clearly (2.3) is invariant under $\NN$ $=$ $1$ supersymmetry. As either version
leads to the same component lagrangian, which is what we need for our
calculations, we will briefly recall its derivation, but from (2.2). This is
obtained by carrying out the integration over the Grassmann coordinates of
(2.2) when each superfield is replaced by its component (or Taylor series)
expansion in powers of $\theta$. The component lagrangian is required as
presently there does not exist a way in the large $N$ critical point approach
for deriving critical exponents in a superspace formulation which would reduce
the amount of calculation, \cite{27}. As the model is a gauge theory one must
choose a gauge in defining the components of $A_\alpha$ which contains the
$U(1)$ gauge field and its supersymmetric partner, the photino. In keeping with
previous work we use the Wess Zumino gauge in which we have, \cite{3,18}, using
the superspace conventions of \cite{27},
\begin{equation}
A_\alpha(x,\theta) ~=~ (\gamma^\mu \theta)_\alpha A_\mu(x) ~+~ (\gamma^5
\theta)_\alpha \pi(x) ~+~ \half \bar{\theta} \theta \omega_\alpha(x)
\end{equation}
where $\pi$ is a bosonic field, $\omega_\alpha$ is a Majorana spinor and
$A_\mu$ is a spin-$1$ field, which we will refer to as the photon. The
components of $\Phi^i$ and $\Lambda$ are defined as in the $O(N)$ $\sigma$
model as
\begin{eqnarray}
\Phi^i(x,\theta) &=& \phi^i(x) ~+~ \bar{\theta}\psi^i(x) ~+~ \half \bar{\theta}
\theta F^i(x) \nonumber \\
\Lambda(x,\theta) &=& \sigma(x) ~+~ \bar{\theta}\delta(x) ~+~
\half \bar{\theta} \theta \lambda(x)
\end{eqnarray}
where $F$ is a bosonic auxiliary field. These then lead to the lagrangian
\begin{eqnarray}
L &=& \partial \bar{\phi} \partial \phi ~+~ i \bar{\psi} \partialslash \psi
{}~+~ iA_\mu \bar{\phi} \stackrel{\leftrightarrow}{\partial^{\mu}}\phi
{}~+~ A_\mu\bar{\psi}\gamma^\mu\psi + \lambda ( \bar{\phi}\phi - 1/g)
\nonumber \\
&+& \bar{u}\psi \bar{\phi} ~+~ \bar{\psi}u\phi ~+~ \sigma \bar{\psi} \psi
{}~+~ \pi \bar{\psi}\gamma^5\psi ~+~ A_\mu A^\mu \bar{\phi}\phi
{}~-~ \sigma^2 \bar{\phi}\phi ~+~ \pi^2\bar{\phi}\phi \label{action}
\end{eqnarray}
where we have eliminated $F$ $=$ $-$ $\sigma\phi$ to obtain the four point
vertex $\sigma^2\bar{\phi}\phi$. We have also combined the auxiliary Majorana
fermion $\delta$ and the Majorana photino $\omega$ into a Dirac fermion $u$,
which now implements the constraint that the complex fermions $\psi^i$ and
$\bar{\psi}^i$ lie in the tangent space to $CP(N)$ which is defined by the
constraint $\bar{\phi}\phi$ $=$ $1/g$, implemented by the Lagrange multiplier
field $\lambda$.

It is worthwhile drawing attention to the field content of the component
lagrangian in dimensions higher than two. For arbitrary dimensions higher than
two, we can still take (2.4) as the component expansion of $A_\alpha$ since
$\gamma^5$ is then defined to be an object which anti-commutes with
$\gamma^\mu$. This implies we are assuming the validity of the $d$-dimensional
$\gamma$-algebra:
\[
\{\gamma_{\mu},\gamma_{\nu}\} ~=~ 2\delta_{\mu\nu}, ~~~~~~
\{\gamma_{\mu},\gamma_{5}\} ~=~ 0, ~~~~~~
\gamma_{5}^{2} ~=~ 1\, ,\,\,\,\gamma_{5}^{\dag} ~=~ \gamma_{5}
\]
This assumption breaks the original two dimensional supersymmetry of the model
which, however, can be restored in three dimensions through the analysis of a
supersymmetric model directly related to (\ref{action}) and obtained from a
standard dimensional reduction procedure explained in the appendix. That three
dimensional model shows that the $\gamma$-algebra must be realized by the Pauli
matrices and consequently no $\gamma_{5}$ appears. So the $\pi$-field is
absent. Bearing this in mind we hope to recover supersymmetry in our analysis
which therefore will be treated in two parts throughout. We will compute
exponents in arbitrary dimensions separately for the lagrangian where there is
a $\pi$-field and where there is not and demonstrate the restoration of
supersymmetry in, say, three dimensions upon specifying to integer dimensions.

One disadvantage in the choice of gauge (2.5) is that it breaks supersymmetry
explicitly. However, the Witten index of the model, \cite{41}, indicates that
supersymmetry is unbroken for models on $CP(N)$. Therefore, provided one
computes gauge independent quantities with the component lagrangian (2.6),
reliable results can be obtained. Indeed this has been used and observed in
other supersymmetric theories in higher dimensions in, for example, \cite{42}.
There the wave function renormalization of the fundamental boson and fermion
fields of the same supermultiplet, which are gauge dependent, are not equal as
one might expect from supersymmetry. However, in computing the
$\beta$-function, say, in either bosonic or fermionic sector, one obtains the
same gauge independent results. The same feature will arise here in relation
to the anomalous dimensions of $\phi^i$ and $\psi^i$, but reliable gauge
independent results will emerge for $\beta(g)$ and for anomalous dimensions of
certain supermultiplets. We qualify further these remarks by noting that for
the $\NN$ $=$ $1$ supersymmetric model on $S^N$ the boson and fermion wave
function renormalization were equal, since the component lagrangian remains
supersymmetric, \cite{33,27}.

Several remarks on gauge fixing are now in order. As we have mentioned (2.6)
possesses a $U(1)$ field $A_\mu$. In the classical theory, it appears as a
bosonic auxiliary field which, in principle, can be eliminated by its equation
of motion as there is no explicit kinetic term like $F^2_{\mu\nu}$. The
presence of such a term would mean that to determine the (perturbative)
$A_\mu$ propagator from (2.6) one would have to introduce a gauge fixing term
first and proceed in a well known formalism. In the bosonic and supersymmetric
models such a term is in fact generated dynamically through particle
anti-particle bubble corrections to the $A_\mu$ field where one computes the
effective action in the true non-perturbative mass generating vacuum,
\cite{2,3,14,15,16}. Moreover, the kinetic term that emerges is in the form
associated with the Landau gauge. Therefore, no conventional gauge fixing is
required, which is the approach used in the large $N$ saddle point technique.
In the critical point formalism of \cite{35}, which we use here, the point of
view is slightly different. There one analyses the theory in the region of the
true vacuum at large energies and uses the form of the propagators which have
the structure consistent with those which would be deduced in the leading
order effective action, say. However, in \cite{35} a gauge dependent $A_\mu$
scaling form for the propagator was employed which was written in terms of a
covariant gauge parameter, which we will call $\rho$ here. Primarily this was
intended to keep track of exponents which were $\rho$-dependent or
$\rho$-independent, with the latter corresponding to physical quantities. For
the supersymmetric model we take the same point of view but do not employ a
supersymmetric gauge fixing term such as
\begin{equation}
\frac{1}{\rho} \int \, d^2x \, d^2 \theta \, (\bar{D} A)D^2 (\bar{D}A)
\end{equation}
There are several reasons for this. First, the component lagrangian (2.6) has
broken supersymmetry anyway and a term such as (2.7) would not alleviate this
difficulty. Second, such a term will lead to the extra terms in the
lagrangian
\begin{equation}
- \, \frac{(\partial_\mu A_\mu)^2}{2\rho} ~+~ \frac{i}{\rho} \bar{\omega}
\partialslash \omega
\end{equation}
In other words, the photino will have a gauge dependent kinetic term (which
would lift the degeneracy with $\delta_\alpha$). However, as the conventional
large $N$ saddle point naturally leads to results in the Landau gauge,
contributions from this photino term would not appear. We ignore this term,
however, and in our calculations take the gauge dependence to be in the photon
field only. This is okay for physical quantities as the gauge dependence
cancels and for our purposes the parameter $\rho$ plays the role of an object
which indicates this explicitly in the calculations.

To proceed with the critical point analysis, and with these remarks in mind, we
now define the asymptotic scaling forms of the fields in the critical region,
$k^2$ $\rightarrow$ $\infty$, in coordinate space. The point of view is that in
the region of the $d$-dimensional fixed point, these scaling functions will be
the dominant form of the propagators, \cite{25}. Thus using the name of each
field of (2.6) as the name of the scaling function, we define the dressed
propagators as
\begin{eqnarray}
\phi(x) &\sim& \frac{A_\phi}{(x^2)^{\alpha_\phi}} ~~~,~~~
\psi(x) ~\sim~ \frac{A_\psi\xslash}{(x^2)^{\alpha_\psi}} \nonumber \\
\lambda(x) &\sim& \frac{B_\lambda}{(x^2)^{\beta_\lambda}} ~~~,~~~
u(x) ~\sim~ \frac{B_u\xslash}{(x^2)^{\beta_u}} \nonumber \\
\sigma(x) &\sim& \frac{B_\sigma}{(x^2)^{\beta_\sigma}} ~~~,~~~
\pi(x) ~\sim~ \frac{B_\pi}{(x^2)^{\beta_\pi}} \nonumber \\
A_{\mu\nu}(x) &\sim& \frac{B_A}{(x^2)^{\beta_A}} \left[ \eta_{\mu\nu} ~+~
\frac{2\beta_A(1-\rho)}{(2\mu-2\beta_A-1+\rho)} \frac{x_\mu x_\nu}{x^2} \right]
\end{eqnarray}
where the spacetime dimension is $d$ $=$ $2\mu$, $A_i$ and $B_i$ are the
$x$-independent amplitudes of the fields and the $\alpha_i$ and $\beta_i$ are
the respective dimensions of the fields. As supersymmetry is broken in (2.6)
we have not assumed that, for example, $\alpha_\phi$ $=$ $\alpha_\psi$ in
keeping with our earlier remarks. Next we define the anomalous piece of each
exponent $\alpha_i$ and $\beta_i$ as
\begin{eqnarray}
\alpha_\phi &=& \mu ~-~ 1 ~+~ \half \eta_\phi ~~~,~~
\alpha_\psi ~=~ \mu ~+~ \half \eta_\psi \nonumber \\
\beta_A &=& 1 ~-~ \eta_\phi ~-~ \chi_{A\phi} ~~,~~
\beta_\sigma ~=~ 1 ~-~ \eta_\psi ~-~ \chi_{\sigma\psi} \nonumber \\
\beta_\pi &=& 1 ~-~ \eta_\psi ~-~ \chi_{\pi\psi} ~~,~~
\beta_u ~=~ 2 ~-~ \half(\eta_\phi ~+~ \eta_\psi) ~-~ \chi_u \nonumber \\
\beta_\lambda &=& 2 ~-~ \eta_\phi ~-~ \chi_\lambda
\end{eqnarray}
where the canonical dimension is given by assuming that the action is a
dimensionless object. In (2.10) the quantities $\chi_i$ correspond to the
anomalous dimension of the relevant vertices of (2.6). Indeed with these it is
possible to deduce various scaling relations between the $\eta$'s and
$\chi$'s just by examining each vertex in (2.6). Thus, for example, the
$A_\mu \bar{\psi}\gamma^\mu\psi$ vertex gives
\begin{equation}
\chi_{A\psi} ~=~ \chi_{A\phi} ~+~ \eta_\phi ~-~ \eta_\psi
\end{equation}
which is an exponent form of a supersymmetric Ward identity in the component
theory. If supersymmetry were unbroken then $\eta_\phi$ would be equivalent
to $\eta_\psi$. Also from the four point term
\begin{equation}
\chi_{\sigma^2\phi^2} ~=~ 2 \eta_\psi ~-~ \eta_\phi ~+~ 2 \chi_{\sigma\psi}
\end{equation}

We now state that our aim is to derive explicit expressions for each of the
anomalous terms of (2.10) at leading order in $1/N$. Each anomalous
dimension is, in $d$-dimensions, a function of $\epsilon$ $=$ $d$ $-$ $2$
and $N$ and can therefore be expanded generally as,
\begin{equation}
\eta(\epsilon,N) ~=~ \frac{\eta_1(\epsilon)}{N} ~+~ \frac{\eta_2(\epsilon)}
{N^2} ~+~ O \left( \frac{1}{N^3} \right)
\end{equation}
and we concentrate on the $\eta_1(\epsilon)$'s. The fixed point is defined to
be the non-trivial zero of the $d$-dimensional $\beta$-function, which here is
\cite{8}
\begin{equation}
\beta(g) ~=~ \epsilon g ~-~ N g^2
\end{equation}
exactly. Hence,
\begin{equation}
g_c ~=~ \frac{\epsilon}{N}
\end{equation}
Another interesting scaling law emerges from the term linear in $\lambda$
which involves the coupling constant. Although $\beta(g)$ vanishes at
criticality one has,
\begin{equation}
- \nu ^{-1} = \beta^\prime(g_c) ~=~ \beta_\lambda ~-~ 2\mu
\end{equation}
Thus to compute the $\beta$-function slope, and from (2.15) the
$\beta$-function to confirm (2.14) in $d$-dimensions, we need only determine
the $\phi$ field anomalous dimension and the $\lambda \bar{\phi}\phi$ vertex
anomalous dimension. As both of these are $\rho$-dependent computing both for
non-zero $\rho$ and observing its cancellation on addition will be an
independent non-trivial check on the result.

As our calculation parallels to a degree the approach of \cite{35} for the
bosonic model, we close this section with the derivation of various fundamental
ingredients which occur in the formalism. The amplitudes $A_i$ and $B_i$ will
arise in computing the Feynman diagrams comprising the Dyson equations we will
solve, in various combinations dictated by the vertices of (2.6). In
particular we define these quantities as follows
\begin{eqnarray}
z &=& A^2_\phi B_\lambda ~~,~~ y ~=~ A^2_\psi B_\sigma ~~,~~ u ~=~ A^2_\phi
B_A \nonumber \\
s &=& A^2_\psi B_\pi ~~,~~ v ~=~ A_\phi A_\psi B_u ~~,~~ t ~=~ A^2_\psi B_A
\end{eqnarray}
As in \cite{35} we now choose to fix two of the amplitudes to be able to
proceed with the analysis. Otherwise the Dyson equations, which are
underdetermined because of gauge invariance, cannot be solved. In the bosonic
case the amplitude of the $\phi$-field in $x$-space was defined by setting its
momentum space form to be $1/(k^2)^{\mu-\alpha_\phi}$. We do so here for $\phi$
as well as demanding that the $\psi$-field is
$\kslash/(k^2)^{\mu-\alpha_\psi+1}$ too. Hence with the Fourier transform
\begin{equation}
\frac{1}{(x^2)^\alpha} ~=~ \frac{a(\alpha)}{2^{2\alpha}\pi^\mu}
\int_k \frac{e^{ikx}}{(k^2)^{\mu-\alpha}}
\end{equation}
and its derivatives and inverse, where $a(\alpha)$ $=$
$\Gamma(\mu-\alpha)/\Gamma(\alpha)$, we set
\begin{equation}
A_\phi ~=~ \frac{a(1)}{4\pi^\mu} ~~~,~~~
A_\psi ~=~ - \, \frac{i(\mu-1)a(1)}{2\pi^\mu}
\end{equation}
Thus one has the useful relation
\begin{equation}
u ~=~ - \frac{t}{4(\mu-1)^2}
\end{equation}

\sect{Computation of anomalous dimensions.}

In this section we detail the calculation of the anomalous dimensions of the
fields and vertices of the component theory as well as constructing a mixing
matrix for a class of operators which have the same quantum numbers. First, we
turn to the Dyson equations for each of the fields of (2.9) and solve them at
leading order in $1/N$ in the critical large $k^2$ region. For compactness,
though, we have illustrated those for the superfields of the $\NN$ $=$ $1$
theory in fig. 1. The Dyson equations for the component fields can readily be
deduced from them. We have omitted the tadpole graphs emanating from the
$4$-point interactions. One could write each of these as a $3$-point vertex by
introducing an auxiliary field like $F$ and $\bar{F}$ and its associated Dyson
equation. However, it turns out that at leading order they give no
contribution. The idea now is to replace the lines in the component graphs by
the critical propagators to represent the critical Dyson equations which can
then be solved, \cite{15}. The scaling forms for the inverse propagators are
required for this which are deduced from (2.9) by inverting their momentum
space form in momentum space before restricting to $x$-space again via (2.18).
For the gauge field one inverts on the transverse subspace in momentum space,
\cite{35,32}. Thus
\begin{eqnarray}
\phi^{-1}(x) &\sim& \frac{p(\alpha_\phi)}{A_\phi(x^2)^{2\mu-\alpha_\phi}}
{}~~~,~~~
\psi^{-1}(x) ~\sim~ \frac{r(\alpha_\psi - 1)\xslash}{A_\psi(x^2)^{2\mu
- \alpha_\psi + 1}} \nonumber \\
\lambda^{-1}(x) &\sim& \frac{p(\beta_\lambda)}{B_\lambda(x^2)^{2\mu
- \beta_\lambda}} ~~~,~~~
\sigma^{-1}(x) ~\sim~ \frac{p(\beta_\sigma)}{B_\sigma(x^2)^{2\mu
- \beta_\sigma}} \nonumber \\
\pi^{-1}(x) &\sim& \frac{p(\beta_\pi)}{B_\pi(x^2)^{2\mu - \beta_\pi}} ~~~,~~~
u^{-1}(x) ~\sim~ \frac{r(\beta_u - 1)\xslash}{B_u(x^2)^{2\mu - \beta_u + 1}}
\nonumber \\
A_{\mu\nu}^{-1}(x) &\sim& \frac{m(\beta_A)}{B_A(x^2)^{2\mu-\beta_A}} \left[
\eta_{\mu\nu} ~-~ \frac{2(2\mu-\beta_A)}{(2\mu-2\beta_A+1)} \frac{x_\mu x_\nu}
{x^2} \right]
\end{eqnarray}
where
\begin{equation}
p(\alpha) ~=~ \frac{a(\alpha-\mu)}{\pi^{2\mu}a(\alpha)} ~~,~~
r(\alpha) ~=~ \frac{\alpha p(\alpha)}{(\mu-\alpha)} ~~,~~
m(\alpha) ~=~ \frac{[4(\mu-\alpha)^2-1] p(\alpha)}{4(\mu-\alpha)^2}
\end{equation}
Hence, one has
\begin{eqnarray}
0 &=& p(\alpha_\phi) ~+~ z ~+~ 2v ~-~ \frac{8(\mu-1)[2(2\mu-1) - \rho\mu]u}
{(2\mu-3+\rho)} \nonumber \\
0 &=& r(\alpha_\psi-1) ~+~ y ~-~ s ~-~ \frac{2[(2\mu-1)(\mu-2)+\rho\mu]t}
{(2\mu-3+\rho)} ~+~ v \nonumber \\
0 &=& p(\beta_\lambda) ~+~ Nz \nonumber \\
0 &=& r(\beta_u-1) ~+~ Nv \\
0 &=& p(\beta_\sigma) ~+~ 2Ny \nonumber \\
0 &=& p(\beta_\pi) ~-~ 2Ns \nonumber \\
0 &=& \frac{(\beta_A-\mu)m(\beta_A)}{(2\beta_A-2\mu-1)} ~+~
\frac{4N(\mu-1)^2u}{(2\mu-1)} ~-~ \frac{2(\mu-1)Nt}{(2\mu-1)} \nonumber
\end{eqnarray}
where we have cancelled off all powers of $x$ at leading order to obtain this
set of homogeneous and algebraic equations. This is valid as these powers only
involve the respective vertex anomalous dimensions, which are $O(1/N)$, and
these therefore do not contribute at leading order. In writing the
representation for $A_\mu$ we have only retained the transverse and therefore
physical part as was implemented in \cite{33,32}. In coordinate space this
corresponds to projecting with $(\eta_{\mu\nu}$ $-$ $2\mu x_\mu x_\nu/x^2)$. It
is not possible to solve (3.3) to find $\eta_\phi$ and $\eta_\psi$ a priori.
Without the condition (2.20), owing to gauge invariance the system is
underdetermined as there would then be seven equations with eight unknowns
which
are $\eta_\phi$, $\eta_\psi$ and the six quantities of (2.17). Therefore as in
the bosonic case, \cite{35}, we have fixed the amplitudes $A_\phi$ and $A_\psi$
as discussed previously. Hence eliminating the amplitude variables, (2.17), and
replacing the $\beta_i$'s by their leading order values defined in (2.10) one
obtains respectively from the boson and fermion sectors
\begin{eqnarray}
\eta_{\phi,1} &=& \frac{\mu(1-\rho)\eta^S_1}{2(\mu-2)} \\
\eta_{\psi,1} &=& - \, \frac{(2\mu^2-3\mu+2+\mu \rho) \eta^S_1}{2(\mu-2)}
\end{eqnarray}
which are clearly not equal. As a non-trivial check, however, one easily
verifies agreement of (3.4) with the same exponent obtained in \cite{36}
within a completely different approach first developed in \cite{37}, where
the $1/N$ expansion technique away from criticality was conjugated with $\MS$
regularization. We will always express all our results in terms of the common
quantity $\eta^S_1$ which is the anomalous dimension of the $\phi$-field in the
bosonic $O(N)$ $\sigma$ model, \cite{25}, and is related to the unit volume of
the $d$-sphere,
\begin{equation}
\eta^S_1 ~=~ - \, \frac{4(\mu-2)\Gamma(2\mu-2)}{\Gamma^2(\mu-1)
\Gamma(\mu+1)\Gamma(2-\mu)}
\end{equation}
In three dimensions $\eta^S_1$ $=$ $8/(3\pi^2)$. In the absence of the
$\pi$-field $\eta_{\phi,1}$ remains unchanged whilst $\eta_{\psi,1}$ becomes
\begin{equation}
\hat{\eta}_{\psi,1} ~=~ - \, \frac{(2\mu^2-4\mu+3+\mu \rho) \eta^S_1}{2(\mu-2)}
\end{equation}
where the hat will denote quantites where $\pi$-field contributions are ignored
in relation to three dimensions. The inequality of these gauge dependent
exponents is illustrative of our earlier remarks on calculating in the
component lagrangian in the Wess Zumino gauge.

We now turn to the computation of the vertex anomalous dimensions, $\chi_i$ at
$O(1/N)$. These, together with a verification of the above results, can be
obtained in the most direct way by carrying out a standard renormalization of
$O(1/N)$ correlation functions with a regularization of dimensional type
extensively explained in \cite{34}, where it is applied to the bosonic $CP(N)$
$\sigma$ model, \cite{35}. As the vertices all are marginal in the $1/N$
expansion for $2$ $\leq$ $d$ $<$ $4$, the standard dimensional regularization
does not work. A regulator $\Delta$ is therefore introduced by a shift in the
dimension of the $\Lambda$ and $A_\alpha$ multiplet fields by $\beta_i$
$\rightarrow$ $\beta_i$ $-$ $\Delta$ which eliminates the marginal character of
all interactions. In practice one uses the propagators of (2.9) with
$\alpha_{\phi}$ $=$ $\mu$ $-$ $1$ $=$ $\alpha_{\psi}$, and the $\beta_i$
replaced by their canonical values. However, unlike (3.3) the vertex Dyson are
considered in momentum space. Necessary for this are the values for the
corresponding amplitudes which are readily deduced from the respective
$x$-space values given by (3.3) after determining (3.4) and (3.5) and using the
Fourier transform, (2.18). Then the graphs contributing to the (bare)
correlation functions show poles in $\Delta$ which will be absorbed by the
renormalization constants through a minimal subtraction. Explicit details of
the application of this in bosonic models can be found in \cite{34,35}. Here we
only remark that a simple relation holds between the renormalization constant
$Z_{i}$ and the anomalous dimension of the $i$-field itself. Writing $Z_{i}$ as
\beq
Z_{i} \ = \ 1 - \frac{\eta _{i}}{2\Delta} \label{Zi} \ ,
\eeq
the relation $G_{\mbox{\small{bare}}}^{2}(i)$ $=$ $Z_{i}
G_{\mbox{\small{ren}}}^{2}(i)$ and the imposition of a scaling form for
$G_{\mbox{\small{ren}}}^{2}(i)$ as in (2.9), easily lead to $\eta_{i}$ $=$
$2\left( \beta_{i}\right.$ $-$
$\left.\beta_{i}^{(\mbox{\small{tree}})}\right.$$\left.\right)$. The subsequent
analysis of the pole part of the vertices, which must be absorbed by the
respective product of the renormalization constants, together with
$\eta_{\phi}$, $\eta_{\psi}$ allow for the determination of the anomalous
dimensions of the remaining fields.

We have summarized the results of this analysis in table 1 where the dotted
line corresponds to the $\psi$-field. This gives the pole part or equally the
contribution from the respective graphs, which we label by Roman numerals, to
the vertex exponent. We have only tabulated the non-vanishing diagrams
necessary to our analysis. The tedious computation was carried out with use of
elementary results for massless Feynman integrals where the exponents of the
momenta in the integral are arbitrary. Also we have included the associated
symmetry factors, such as minus signs for fermion loops, as well as the
momentum space values of the quantities corresponding to (2.17). First, from
the sum of (I-III) and (IV, V) one deduces that $\beta_{A}$ $=$ $2$ $+$
$O(1/N^2)$, or in other words that $\chi_{A\phi,1}$ $=$ $-$ $\eta_{\phi,1}$,
$\chi_{A\psi,1}$ $=$ $-$ $\eta_{\psi,1}$ which is a consequence of gauge
invariance. Another way of putting this is that the structure of our scheme
preserves the ($U(1)$) gauge invariance so that $\beta_{A}$ $=$ $2$ follows
directly from a Ward identity. Moreover, this result can also be considered as
an independent check of (3.4) and (3.5). Next in the same way, diagrams
(VI,VII), (VIII-XI), (XII-XVI) and relations (2.10) yield respectively:
\begin{eqnarray}
\eta_{\sigma} &=& \eta _{\pi} = \ \frac{2(3\mu - 1)}{N(2 - \mu )}\
\eta_{1}^{S}
\label{etas}\\
\eta _{u} &=& - \frac{(4\mu - 1)(\mu - 1)^{2}}{N(\mu - 2)}\ \eta _{1}^{S}
\label{etau}\\
\eta _{\lambda} &=& 0 \label{etad}
\end{eqnarray}
where we have defined $\eta_\sigma$ etc in the same way as, for example,
$\eta_\phi$ of (2.10). Besides the disappearance of the parameter $\rho$ in
these exponents corresponding to gauge ($U(1)$) independent fields, we stress
that from (3.11), which is strictly valid only for $\mbox{tr}1$ $=$ $2$, the
critical index $\nu$ takes the value, (2.16),
\beq
\nu^{-1} ~=~ - \beta^\prime (g_c) ~=~ 2\mu ~-~ 2 \label{nu1}
\eeq
So the absence of corrections to the $\beta$-function of the supersymmetric
$CP(N)$ $\sigma$ beyond one loop is extended at $O(1/N)$ for any $2$ $\leq$ $d$
$<$ $4$. Nevertheless we must take care when analyzing an extension of the
model in $d$ $\neq$ $2$, as our $\gamma$-algebra rules are sensible only at a
formal level and it is not obvious which properties of the original
supersymmetric theory survive in $d$ $\neq$ $2$.

A possible alternative way of computing (3.12) is by considering the anomalous
dimension of the coupling associated with the operator $\sigma^2$. For example,
in the $O(N)$ Gross Neveu model there is a term $\sigma^2/2g$ in the
lagrangian. So if one considers the insertion of the composite operator
$\sigma^2$ in a Green's function to deduce $\eta_{\sigma^2}$, using the method
discussed in \cite{34,35}, then one correctly recovers the $O(1/N)$ correction
to $\beta^\prime(g_c)$ there. To investigate this in the supersymmetric $CP(N)$
$\sigma$ model we analyse the scale dimension of the analogous operator
$(\sigma^2 - \pi^2)$ which appears in the lagrangian (2.6) after the shift
\begin{equation}
\lambda(x) \longrightarrow \lambda(x) + \sigma^{2}(x) - \pi^{2}(x)
\end{equation}
or, equivalently, by using the constraint $\bar{\phi} \phi$ $=$ $1/g$ to
eliminate the $4$-point interactions. By counting the dimensions of the fields,
however, this operator can mix with several others. Therefore we will consider
the following set of independent operators of dimension $2$,
\begin{equation}
Q_i(x) = \left[ \lambda, \ \half \sigma^{2}, \ \half \pi^{2}, \
\half A^{\mu}A_{\mu} \ \right]\ .
\end{equation}
where $1$ $\leq$ $i$ $\leq$ $4$. The calculation of the mixing renormalization
constants rests on the analysis of the amplitudes:
\beq
\langle Q_{k}(x) \sigma (y) \sigma(z) \rangle^{1PI}\ \ \ \ \ \langle Q_{k}(x)
\pi (y) \pi (z) \rangle ^{1PI}\ \ \ \ \
\langle Q_{k}(x) \lambda(y) \rangle ^{1PI}\ \ \ \ \ \langle Q_{k}(x) A_{\mu}(y)
A_{\nu}(z) \rangle ^{1PI}
\eeq
where the superscript signifies amputation of the external lines except
$\lambda$. One can argue almost directly \cite{35} that $\langle Q_{k} \lambda
\rangle$ $=$ $0$ if $Q_{k}$ $\neq$ $\lambda$ from multiplicative
renormalizability, $\langle Q_{k} v_{\mu} v_{\nu} \rangle$ $=$ $0$ if $ Q_{k}$
$\neq$ $A^{2}/2$ from transversality and logarithmic divergence and finally
that $\langle \sigma^{2} \sigma \sigma \rangle$ $=$ $\langle \pi^{2} \pi \pi
\rangle$, $\langle \sigma^{2} \pi \pi \rangle$ $=$ $\langle \pi^{2} \sigma
\sigma \rangle$ from chiral invariance. (Strictly speaking this follows from
the assumption of the existence of an anticommuting $\gamma^{5}$.) As we are
considering only the leading order, $O(1/N)$, the preceding amplitudes
entirely determine the matrix elements. Indeed if we let
$\triangleleft \ X \ \triangleright$ mean the pole part of $X$, we obtain the
terms we are interested in, like $\triangleleft Z_{Q_{k},\sigma^{2}}
\triangleright$, from
\beq
\triangleleft \langle Q_{k}\sigma \sigma \rangle \triangleright ~=~
\triangleleft Z_{Q_{k},\sigma^{2}}Z_{\sigma}^{-1} \langle \sigma^{2} \sigma
\sigma \rangle_{\mbox{\small{ren}}}\triangleright ~+~
O(1/N^{2}) ~=~ \triangleleft Z_{Q_{k},\sigma^{2}}Z_{\sigma}^{-1} \triangleright
\eeq
because the subleading structure of the other couplings yield, for example,
\beq
\triangleleft Z_{Q_{k},\pi^{2}}Z_{\sigma}^{-1} \langle \pi ^{2}\sigma \sigma
\rangle_{\mbox{\small{ren}}} \triangleright ~=~ O\left(\frac{1}{N^{2}}\right)
\eeq
and so on. Consequently, summing the poles (XVIIa, XVIII) we obtain:
\beq
\triangleleft Z_{\sigma^{2},\sigma^{2}}Z_{\sigma}^{-1} \triangleright ~=~
\triangleleft Z_{\pi^{2},\pi^{2}}Z_{\pi}^{-1} \triangleright ~=~
\frac{\mu (2\mu - 1)}{2(\mu - 2)N}\eta _{1}^{S}
\eeq
and from (XVIIb, XVIII):
\beq
\triangleleft Z_{\sigma^{2},\pi^{2}}Z_{\pi}^{-1} \triangleright ~=~
\triangleleft Z_{\pi^{2},\sigma^{2}}Z_{\sigma}^{-1}
\triangleright ~=~ - \, \frac{\mu (2\mu - 3)}{2(\mu - 2)N}\eta _{1}^{S}
\eeq
Now setting $Z_{a,b}$ $=$ $1$ $-$ $\frac{\eta_{a,b}}{2\Delta}$ and making use
of (\ref{etas})-(\ref{etad}) and the above remarks on a priori zero amplitudes,
one obtains:
\beq
\eta _{\sigma^{2}, \sigma^{2}} ~=~ \eta _{\pi^{2}, \pi^{2}} ~=~
\frac{2(2\mu^{2} + 5\mu - 2)}{(2 - \mu)N}\ \eta_{1}^{S}\ \ \ \ \ \ \ \ \
\eta_{\sigma^{2}, \pi^{2}} ~=~ \eta_{\pi^{2}, \sigma^{2}} ~=~
\frac{2(2\mu ^{2} - 9\mu + 2)}{(\mu - 2)N}\ \eta_{1}^{S}
\eeq
\beq
\eta_{\lambda, Q_{k}} ~=~ \eta_{\sigma^{2}, \lambda} ~=~
\eta_{\sigma^{2}, A^{2}} ~=~
\eta_{\pi^{2}, \lambda} ~=~ \eta_{\pi^{2}, A^{2}} ~=~ 0
\eeq
We do not need an explicit evaluation of $\eta_{A^{2},Q_{k}}$, because from the
structure of the matrix we see that
\beq
\eta_{\sigma^{2}, \sigma^{2}} - \eta_{\sigma^{2},\pi^{2}} ~=~
\eta_{[\sigma^{2}-\pi^{2}]} ~=~ - \frac{8\mu (\mu - 1)}{N(\mu-2)}\eta_{1}^{S}
\eeq
is the eigenvalue corresponding to the expected eigenoperator $(\sigma^{2} -
\pi^{2})$. Clearly (3.22) does not vanish, although (3.18) correctly gives the
Gross Neveu $\beta$-function slope. The reason for this apparent discrepancy
lies in our initial use of the constraint to modify the lagrangian. Such a step
breaks supersymmetry explicitly. The lagrangian cannot be written in a
manifestly supersymmetric form using, for example, superfields. Indeed this
calculation of the $\sigma^2$ scaling dimension and its non-zero value is an
indication of some of the perils one would encounter by using the constraint to
modify (2.6), which unfortunately is used as a simplifying procedure in the
literature.

In the appendix we show how a four dimensional supersymmetric model can be
dimensionally reduced to obtain our $CP(N)$ supersymmetric $\sigma$ model in
$d$ $=$ $2$ or an equivalent one at the critical point. The same reduction
leads to the following euclidean action when $d$ $=$ $3$
\beq
S_{E} ~=~ \int d^{3}x\;\left[D^{\mu}\bar{\phi} D_{\mu}\phi
+ i\bar{\psi}\,{\Dslash}\psi - \sigma^{2}\bar{\phi}\phi + \sigma\bar{\psi}\psi
- \bar{F}F \right] - \left[ \lambda \bar{\phi}\phi + \bar{\psi}u \phi
+ \bar{u} \psi\bar{\phi}\right] + \ldots
\eeq
where $\gamma_{\mu}$ $=$ $(\vec{\sigma})$ and $F$ is an auxiliary field. No
$\pi$-field is present because no $\gamma^{5}$ arises when the Clifford algebra
is realized through the Pauli matrices. The previous Schwinger Dyson equations
and the diagrammatics can still be used provided that: (a) the terms in (3.1)
and the graphs in the table with $\pi$-propagators are ignored, (b) the
additional graph in fig. 2 is taken into account in the analysis of the
$\sigma\bar{\psi}\psi$ vertex. It does not vanish because now
$\mbox{tr}(\gamma_{\mu}\gamma_{\nu}\gamma_{\rho})$ $=$
$2i\epsilon_{\mu\nu\rho}$ where $\epsilon_{\mu\nu\sigma}$ is totally
antisymmetric in three dimensions. We get the following values for the
anomalous dimensions, in addition to (3.7),
\begin{eqnarray}
\hat{\eta}_{\phi,1} ~+~ \hat{\chi}_{\lambda,1} &=& 0 \nonumber \\
\hat{\eta}_{\psi,1} ~+~ \hat{\chi}_{\sigma\psi,1} &=& - \,
\frac{(2\mu-3)(2\mu-1)\eta^S_1}{2(\mu-2)} \nonumber \\
\half(\hat{\eta}_{\phi,1} + \hat{\eta}_{\psi,1}) ~+~ \hat{\chi}_{u,1}
&=& \frac{(4\mu-1)(2\mu-3)(\mu-1)\eta^S_1}{4(\mu-2)} \\
\hat{\eta}_{\phi,1} ~+~ \hat{\chi}_{A\phi,1} &=& 0 \nonumber \\
\hat{\eta}_{\psi,1} ~+~ \hat{\chi}_{A\psi,1} &=& 0 \nonumber
\end{eqnarray}
We have expressed our results in $d$-dimensions to allow the interested reader
an intermediate check on assemblying the contributions from the various graphs
of the table, in the same vein as (3.9)-(3.11), and to illustrate several
features. For example, the $U(1)$ Ward identity is still evident in the final
two results. Moreover, though we have obtained the result of (3.12) in the
first equation, it is unaffected anyway by neglecting the $\pi$-field. If
one analyses the graphs making up both the relevant $2$-point Dyson equations
and vertex graphs, there are no graphs involving a $\pi$-field. Strictly these
expressions are only valid in integer dimensions and the restoration of
supersymmetry in three dimensions can be easily determined as a consequence of
the explicit factors of $(2\mu$ $-$ $3)$.

\sect{Chern Simons term.}
In this section we discuss the effect the inclusion of a Chern Simons term,
\cite{38}, has on our results in strictly three dimensions. Such a term arises,
of course, from dimensionally reducing \cite{38}, $\epsilon_{\mu\nu\sigma\rho}
F^{\mu\nu} F^{\sigma\rho}$ from four dimensions to three. The aim is to
determine whether the $O(1/N)$ correction to $\beta^\prime(g_c)$ remains zero
or whether it depends on the Chern Simons coupling, $\vartheta$, which is a
continuous dimensionless parameter for an abelian theory which does not get
renormalized, \cite{39}. In the bosonic model such corrections have also been
calculated recently and $\vartheta$-dependent exponents emerged, but there the
critical slope at $O(1/N)$ was non-zero for $\vartheta$ $\neq$ $0$, \cite{40}.
Whilst those large $N$ calculations used the conventional saddle point
approach, that method becomes tedious in the supersymmetric case compared to
the more efficient critical point analysis.

The inclusion of a Chern Simons term in $\NN$ $=$ $2$ supersymmetric theories
has been considered in other contexts before, \cite{45,46,47}. In \cite{46} the
motivation was to examine the fixed point structure of abelian and
non-abelian theories for a variety of couplings and examine their stability.
Also the dynamics of topological objects has been explored in \cite{47} in
the abelian Higgs model with extended supersymmetry. As $CP(N)$ is an abelian
theory it admits the possibility of including the term $\vartheta
\epsilon_{\mu\nu\sigma}A^\mu \partial^\nu A^\sigma$. (We use the conventions of
\cite{48} for defining the normalization of $\vartheta$.) As we are in a
supersymmetric theory such an additional term must be included in the action
in a supersymmetric fashion. For example, the term, \cite{47},
\begin{equation}
\vartheta \, \int d^3x \, d^2\theta \,  \bar{D}^\alpha \bar{A}^\beta D_\beta
A_\alpha
\end{equation}
in the $\NN$ $=$ $1$ formulation yields the additional terms in (2.6),
\begin{equation}
\vartheta [ \epsilon_{\mu\nu\sigma}A^\mu \partial^\nu A^\sigma ~+~
\quarter i \bar{\omega} \omega ]
\end{equation}
which contains the Chern Simons term. With (4.1), however, the full model will
only be invariant under one supersymmetry and not two. Despite this we will
first outline the technicalities involved in the calculation of our earlier
results with (4.2) as a prelude to considering a Chern Simons term which is
supersymmetric under the full $\NN$ $=$ $2$. So with (4.2) in addition to the
usual Chern Simons term there is an extra $\vartheta$ dependent contribution to
the Majorana photino field. Its presence lifts the degeneracy with the Majorana
field, $\delta$, and so we must treat these fields separately now for
$\vartheta$ $\neq$ $0$. (This situation may be construed to be similar to the
argument concerning the gauge fixing. Unlike that case we cannot omit the
second term of (4.2).) As both terms of (4.2) are quadratic this leads to new
propagators which must be used in the critical analysis. For example, the
photon field scaling form is now, in $x$-space,
\begin{equation}
A_{\mu\nu}(x) ~\sim~ \frac{B_A}{(x^2)^{\beta_A}} \left[ \eta_{\mu\nu} ~+~
\frac{2\beta_A(1-\rho)}{(2-2\beta_A+\rho)} \frac{x_\mu x_\nu}{x^2} ~+~
i \vartheta h(\beta_A) \frac{\epsilon_{\mu\nu\sigma} x^\sigma}{(x^2)^{1/2}}
\right]
\end{equation}
where
\begin{equation}
h(\beta_A) ~=~ \frac{2(3-2\beta_A) a(1-\beta_A)}{(1-\beta_A)^2 a(\threehalves
- \beta_A)}
\end{equation}
or in momentum space,
\begin{equation}
\tilde{A}_{\mu\nu}(k) ~\sim~ \frac{\tilde{B}_A}{(k^2)^{3/2-\beta_A}}
\left[ \eta_{\mu\nu} ~-~ (1-\rho)\frac{k_\mu k_\nu}{k^2} ~+~ \vartheta
\frac{\epsilon_{\mu\nu\sigma}k^\sigma}{(k^2)^{1/2}} \right]
\end{equation}
where the $~\tilde{}~$ denotes quantities defined in momentum space and, for
example, $\tilde{B}_A$ is related to $B_A$ by a factor deduced from (2.18). We
note that this form differs from the Chern Simons terms which are dynamically
generated and which have been studied for (2.6) in, for example, \cite{49}. In
that case the parameter $\vartheta$ involves the mass of the fundamental
particle supermultiplet. As an aside we note that we have checked the exponent
results of \cite{40} for the bosonic model in the critical point approach using
(4.3) and find exact agreement. Consequently inverting (4.3) on the transverse
subspace to obtain the scaling form of the inverse propagator leads to a
structure similar to $\vartheta$ $\neq$ $0$, but now its amplitude is
explicitly $\vartheta$-dependent, \cite{48},
\begin{equation}
\tilde{A}^{-1}_{\mu\nu}(k) ~\sim~ \frac{1}{\tilde{B}_A(1+\vartheta^2)
(k^2)^{\beta_A-3/2}} \left[ \eta_{\mu\nu} ~-~ \frac{k_\mu k_\nu}{k^2}
{}~-~ \vartheta \frac{\epsilon_{\mu\nu\sigma}k^\sigma}{(k^2)^{1/2}}
\right]
\end{equation}
Likewise the photino propagator, $\omega_\alpha$, is now
\begin{equation}
\tilde{\omega}(k) ~\sim~ \tilde{B}_\omega \left[ \frac{\kslash}{(k^2)^{1/2}}
{}~+~ i \vartheta \right]
\end{equation}
whence
\begin{equation}
\tilde{\omega}^{-1}(k) ~\sim~ \frac{1}{(1+\vartheta^2)\tilde{B}_\omega}
\left[ \frac{\kslash}{(k^2)^{1/2}} ~-~ i \vartheta \right]
\end{equation}
which also has an explicitly $\vartheta$-dependent amplitude. Moreover, in
doing calculations one must now treat the contributions from the Majorana
fields $\delta$ and $\omega$ distinctly.

There are several consequences of these new propagators which simplify the
determination of exponents. First, each time a photon or photino line appears
in a graph for $\vartheta$ $=$ $0$ one now divides by $(1+\vartheta^2)$ per
line. Second the results will depend on $\vartheta^2$ and not $\vartheta$.
This follows directly from the fact that the presence of an $\epsilon$-tensor
on its own in an integral gives zero since it is always contracted with at
least two equal momenta. Therefore the only place additional $\vartheta$
dependence can arise in numerators of exponents is in graphs with a
combination of two photon or photino lines, when the identity
\begin{equation}
\epsilon_{\mu\nu\sigma} \epsilon^\sigma_{~\lambda\rho} ~=~ \eta_{\mu\lambda}
\eta_{\nu\rho} ~-~ \eta_{\mu\rho}\eta_{\nu\lambda}
\end{equation}
is used to give a contribution. Thus, for example,
\begin{eqnarray}
&& \left[ \eta_{\mu\nu} - (1-\rho)\frac{k_\mu k_\nu}{k^2} + \vartheta
\frac{\epsilon_{\mu\nu\rho}k^\rho}{(k^2)^{1/2}} \right]
\left[ \eta^{\nu\sigma} - (1-\rho)\frac{k^\nu k^\sigma}{k^2} - \vartheta
\frac{\epsilon^{\nu\sigma\lambda}k_\lambda}{(k^2)^{1/2}} \right] \nonumber \\
&&=~ (1+\vartheta^2)\eta_\mu^{~\sigma} ~-~ (1-\rho^2+\vartheta^2)
\frac{k_\mu k^\sigma}{k^2}
\end{eqnarray}
and
\begin{equation}
\left[ \frac{\kslash}{(k^2)^{1/2}} ~+~ i \vartheta \right]
\left[ \frac{\kslash}{(k^2)^{1/2}} ~-~ i \vartheta \right] ~=~ 1 ~+~
\vartheta^2
\end{equation}
where in explicit calculations one must be careful to choose the momentum flow
correctly.

With these intermediate observations it is straightforward to return to the
arbitrary dimensional calculation of $\eta_\phi$ and $\chi_{\lambda\phi}$ and
evaluate the result for each graph in three dimensions and to include the
$\vartheta$ dependence. (Throughout this calculation we have retained the
convention that $\mbox{tr}1$ $=$ $2$.) Therefore we have in long form, where
each term within the bracket corresponds to a contributing Feynman graph in
(3.3), that
\begin{equation}
\hat{\eta}_{\phi,1} ~=~ \frac{8}{3\pi^2}\left[ \frac{1}{2} + 1
- \frac{(8-3\rho)}{2(1+\vartheta^2)} + \frac{1}{(1+\vartheta^2)} \right]
\end{equation}
to yield the simple result
\begin{equation}
\hat{\eta}_{\phi,1} ~=~ \frac{4(\rho-1+\vartheta^2)}{\pi^2(1+\vartheta^2)}
\end{equation}
Likewise the computation of the vertex anomalous dimension can be summarized
in long form as
\begin{equation}
\hat{\chi}_{\lambda\phi,1} ~=~ - \, \frac{4}{\pi^2}\left[ - \, 1
+ \frac{\rho}{(1+\vartheta^2)} - \frac{(1-\vartheta^2)}{(1+\vartheta^2)^2} + 2
+ \frac{2(1-\vartheta^2)}{(1+\vartheta^2)^2} \right]
\end{equation}
In three dimensions the contribution of one graph is zero. The last two terms
of (4.14) correspond to contributions involving $\delta_\alpha$ and
$\omega_\alpha$. Now that their propagators are different we have split their
contributions appropriately. Thus
\begin{equation}
\hat{\chi}_{\lambda\phi,1} ~=~ - \, \frac{4}{\pi^2(1+\vartheta^2)^2}
[\vartheta^4 + 4\vartheta^2 - 1 + \rho (1+\vartheta^2)]
\end{equation}
Now having established in section 3 that the use of the scaling law (2.16)
determines $\beta^\prime(g_c)$ correctly, we again use that approach to
deduce the correction to the critical $\beta$-function slope is given by
\begin{equation}
\hat{\eta}_{\phi,1} ~+~ \hat{\chi}_{\lambda\phi,1} ~=~
- \, \frac{16\vartheta^2}{\pi^2 (1+\vartheta^2)^2}
\end{equation}
which is clearly non-zero. We have also repeated the calculation of the field
exponents and find
\begin{eqnarray}
\hat{\eta}_{\psi,1} ~+~ \hat{\chi}_{\sigma\psi,1} &=& \half
(\hat{\eta}_{\phi,1} + \hat{\eta}_{\psi,1}) ~+~ \hat{\chi}_{\delta,1} ~=~
- \, \frac{16\vartheta^2}{\pi^2(1+\vartheta^2)^2} \nonumber \\
\hat{\eta}_{\phi,1} ~+~ \hat{\chi}_{A\phi,1}
&=& \hat{\eta}_{\psi,1} ~+~ \hat{\chi}_{A\psi,1} ~=~ \half
(\hat{\eta}_{\phi,1} + \hat{\eta}_{\psi,1}) ~+~ \hat{\chi}_{\omega,1} ~=~ 0
\end{eqnarray}
where $\hat{\chi}_\delta$ and $\hat{\chi}_u$ are the vertex anomalous
dimensions of the respective Majorana fields defined in (2.4) and (2.5) which
also have to be treated separately. Also in considering the $\sigma \bar{\psi}
\psi$ vertex we have been careful to include the contribution from the extra
graph of fig. 2 which occurs precisely in three dimensions.

Several comments on these results are in order. Clearly (4.16) and (4.17) are
consistent with the $\NN$ $=$ $1$ supersymmetry which results from including
(4.1) in the lagrangian. In other words the fields of each $\Lambda$ and
$A_\alpha$ supermultiplets have the same anomalous dimensions. Second, one
trivially recovers the earlier results as $\vartheta$ $\rightarrow$ $0$. More
substantially one can examine the $\vartheta$ $\rightarrow$ $\infty$ behaviour.
Returning to the action (2.6) with the term (4.2) this limit clearly
corresponds to integrating out $A_\mu$ or equivalently setting it and its
superpartner, $\omega_\alpha$, to zero. Therefore one is left with the
supersymmetric $\sigma$ model on the $2N$-sphere. So as a non-trivial check we
ought to recover the critical exponents for that model as $\vartheta$
$\rightarrow$ $\infty$. These were computed in exponent form in \cite{27,33}
and we note that for $S^N$ in three dimensions
\begin{equation}
\eta_\phi ~=~ \frac{8}{\pi^2N} ~+~ O\left( \frac{1}{N^2} \right) ~~~,~~~
\chi_{\lambda\phi} ~=~ - \,  \frac{8}{\pi^2N} ~+~ O\left( \frac{1}{N^2} \right)
\end{equation}
and $\beta^\prime(g_c)$ $=$ $-$ $1$ $+$ $O(1/N^2)$. It is easy to verify these
results are correctly recovered. Further (4.16) is curiously unchanged under
the transformation $\vartheta$ $\rightarrow$ $1/\vartheta$. The analogous
result in related models does not share this feature. For example, in the
bosonic model, \cite{40},
\begin{equation}
\eta_{\phi,1}^{\mbox{\small{bos}}}
{}~+~ \chi_{\lambda\phi,1}^{\mbox{\small{bos}}} ~=~
\frac{16(\vartheta^4 - 14\vartheta^2 + 9)}{3\pi^2(1+\vartheta^2)^2}
\end{equation}
Indeed its functional form is quite different in that it is negative only
for $7$ $-$ $2\sqrt{10}$ $<$ $\vartheta^2$ $<$ $7$ $+$ $2\sqrt{10}$. Further
repeating the same calculation for the minimal extension of the $CP(N)$ model,
\cite{31}, there
\begin{equation}
\eta_{\phi,1}^{\mbox{\small{min}}}
{}~+~ \chi_{\lambda\phi,1}^{\mbox{\small{min}}} ~=~
\frac{16(1-\vartheta^2)(2 - \vartheta^2)}{3\pi^2(1+\vartheta^2)^2}
\end{equation}
which is negative for $1$ $<$ $\vartheta^2$ $<$ $2$. Indeed it has a similar
structure to (4.19) though the addition of fermions forces the zeroes to be at
integer values. Moreover, (4.20) tends to the bosonic value as $\vartheta$
$\rightarrow$ $\infty$ which is consistent with the decoupling of the fermions.
Thus this invariance appears to follow directly from the presence of
supersymmetry in a similar way that in the $\vartheta$ $=$ $0$ sector
supersymmetry implies there are no $O(1/N)$ contributions.

Finally, we consider the model where an $\NN$ $=$ $2$ Chern Simons term is
included. In terms of vector superfields this is, \cite{47},
\begin{equation}
\vartheta \, \int \, d^3x d^2 \bar{\theta} d^2\theta \, \bar{D}V D V
\end{equation}
Alternatively in component language we have the additional term
\begin{equation}
\vartheta [ \epsilon_{\mu\nu\sigma}A^\mu \partial^\nu A^\sigma ~+~
\quarter i \bar{u}u + \lambda \sigma ]
\end{equation}
Here the Dirac fermion appears in contrast to the $\omega$ of (4.2). Also there
is now a $\lambda\sigma$ mixing term. With these extra terms the propagators of
the fields are modified, though that of $A_\mu$ remains as (4.5). It is now the
$u$-field whose propagator has the form (4.7) and there is no need to treat
$\omega$ and $\delta$ distinctly. The only difficulty lies in the form of the
$\lambda$ and $\sigma$ fields. Due to the mixing term one could, for instance,
diagonalize the $2$ $\times$ $2$ matrix of propagators to determine the form of
the propagators of the eigenfields and use these to perform the calculation.
This turns out to be tedious and it is simpler, and equivalent, to use a mixed
two point function $\langle \lambda \sigma \rangle$ and $\langle \sigma \lambda
\rangle$ in the Feynman diagrams. Therefore we take as the form for the $2$
$\times$ $2$ matrix, with respect to the basis $(\lambda,\sigma)$,
\begin{equation}
\left(
\begin{array}{cc}
\frac{\tilde{B}_\lambda}{(k^2)^{3/2-\beta_\lambda}} &
i\vartheta \frac{\tilde{B}_{\lambda\sigma}}{(k^2)^{\beta_{\lambda\sigma}}} \\
i\vartheta \frac{\tilde{B}_{\lambda\sigma}}{(k^2)^{\beta_{\lambda\sigma}}}
& \frac{\tilde{B}_\sigma}{(k^2)^{3/2-\beta_\sigma}} \\
\end{array}
\right)
\end{equation}
in $k$-space, where the mixed propagator has zero canonical dimension,
$\beta_{\lambda\sigma}$ $=$ $\eta_{\lambda\sigma}$. We need to find
$\tilde{B}_{\lambda\sigma}$ in relation to the other amplitudes. This is
obtained from the determinant of (4.6) which has to be proportional to the
factor $(1+\vartheta^2)$ which arises in from the analogous derivation of the
same factor in (4.6), in order to have unbroken supersymmetry. (We have checked
that this factor emerges naturally in a superspace calculation.) As the
determinant is $[\tilde{B}_\lambda \tilde{B}_\sigma$ $+$ $\vartheta^2
\tilde{B}^2_{\lambda\sigma}]$, omiting the momentum dependence which factors
out at leading order in $1/N$, then
\begin{equation}
\tilde{B}_{\lambda\sigma} ~=~ \sqrt{\tilde{B}_\lambda \tilde{B}_\sigma}
\end{equation}
Using the explicit values of the amplitudes the mixed amplitude is
$a^2(\mu-1)/a(2\mu-2)$. The upshot of this is that in addition to the earlier
Chern Simons Feynman rules we had for the $\NN$ $=$ $1$ case, we must now
divide all contributions with $\lambda$ and $\sigma$ fields by
$(1+\vartheta^2)$ and also compute several extra graphs which arise involving
the mixed propagator and which are easy to determine.

Therefore reanalysing our earlier work we note that no new graphs occur for the
exponents $\eta_\phi$ and $\eta_\psi$ which now are, for the full $\NN$ $=$
$2$,
\begin{equation}
\hat{\eta}_\phi ~=~ \frac{4(\rho-1)}{\pi^2(1+\vartheta^2)} ~~~,~~~
\hat{\eta}_\psi ~=~ \frac{4(\rho+1)}{\pi^2(1+\vartheta^2)}
\end{equation}
Further, it transpires that the exponents of all the other fields vanish
\begin{equation}
\hat{\eta}_\lambda ~=~ \hat{\eta}_\sigma ~=~ \hat{\eta}_u ~=~ \hat{\eta}_A ~=~
0
\end{equation}
There are several checks now on these results. First, the anomalous dimensions
of the fields $\lambda$, $\sigma$ and $u$, being gauge invariant, are equal to
one another as they ought to be with regard to the choice of the Wess Zumino
gauge. Moreover, they vanish together with $\eta_{A}$, and this respects the
vector multiplet structure and reestablishes $\NN$ $=$ $2$ supersymmetry. If
any of (4.26) had been different then this would have been broken. Second, the
fact that they are zero implies that for $\NN$ $=$ $2$ Chern Simons the
correction to the critical $\beta$-function slope is the same as in the absence
of such a term. A final check is that in the $\vartheta$ $\rightarrow$ $\infty$
limit the vector supermultiplet decouples completely from the lagrangian to
leave a free field lagrangian and it is trivial to observe that
$\hat{\eta}_\phi$ and $\hat{\eta}_\psi$ both correctly tend to zero in this
limit.

\sect{Conclusions.}
We have now completed the leading order $1/N$ analysis of the supersymmetric
$CP(N)$ $\sigma$ model using the critical exponent technique. In particular we
have verified the $\beta$-function slope at criticality gains no $O(1/N)$
corrections both in two dimensions, which is consistent with other results,
and now also in three dimensions. Also in the latter dimension we have
demonstrated that supersymmetry is restored, as one varies the spacetime
dimension. Whilst it may appear disappointing that there are no $O(1/N)$
corrections to any of the anomalous dimensions, this is in fact a consequence
of the many symmetries present in the model. Indeed a similar feature was
observed in the supersymmetric $O(N)$ $\sigma$ model, \cite{27}. In that case,
however, it turned out that there were non-zero corrections at $O(1/N^2)$
which did not vanish in three dimensions. For the $CP(N)$ case, it is tempting
to speculate on what may lie beyond $O(1/N)$ in light of our calculations. For
instance, the absence of corrections beyond $O(g^3)$ in (2.14), which we have
verified in (3.13), suggests that the $\Lambda$-multiplet will have zero
anomalous dimension at all orders in $1/N$ in $d$-dimensions, if one assumes
supersymmetry is unbroken in the model. Moreover, the gauge multiplet anomalous
dimension are also zero at $O(1/N)$, a feature which is similar to what occurs
in QED, \cite{32}. In that model, in the way that the critical point analysis
is carried out, the $A_\mu$ field has no anomalous dimension to all orders due
to the Ward identity. If such an identity holds in the supersymmetric $CP(N)$
$\sigma$ model then the anomalous dimension of that multiplet would vanish too.
(The non-vanishing of $\eta_\phi$ and $\eta_\psi$ is unimportant here as they
are gauge dependent and therefore not meaningful.) Of course to justify such a
point of view further, an $O(1/N^2)$ computation would need to be performed,
which would be tedious though the technology does exist to do this now. Indeed
this paper would be a first stage in that direction.

In studying the effect a Chern Simons term has, we have observed differing
behaviour in $\beta^\prime(g_c)$ depending on the number of supersymmetries the
additional term is invariant under. In the case of $\NN$ $=$ $1$ supersymmetry
we would not expect the $\vartheta$ $\rightarrow$ $1/\vartheta$ invariance of
the $O(1/N)$ exponent to be a feature of the full result. Although the former
will yield zero through the observation (2.14), the latter limit will be
non-zero at $O(1/N^2)$. This is because one must recover the exponents of the
$2N$-sphere which has a non-zero $O(1/N^2)$ contribution to
$\beta^\prime(g_c)$, \cite{27}, which must remain after taking the limit.
Finally, aside from calculating $O(1/N^2)$ corrections to our results it would
be interesting to pursue an examination of the current algebra of the
supersymmetric model along the lines of \cite{40} for the bosonic model to try
and gain a further insight into the fixed point structure of the model.

\vspace{1cm}
\noindent
{\bf Acknowledgement.} One of us (M.C.) wishes to thank Prof. P. Rossi for
useful conversations.

\appendix

\sect{Dimensional reduction.}

In this section we justify the use of the three dimensional lagrangian we used
in section 3 from dimensional reduction. This method was introduced in
\cite{50,51}. Here we recall only the essential steps of the procedure. The
starting point is to consider one or more dimensions to be compact and then to
let its measure shrink to zero. In practice one introduces a radius $R_{i}$
such that $0$ $\leq$ $x_{i}$ $\leq R_{i}$ and, after a rescaling, only the
zeroth coefficient in the Fourier expansion of the fields is retained so that
the analysis is focused on the effects of the $\gamma$-algebra. In our view the
simplest way to derive our results concerning the absence of the $\pi$-field in
three dimensions is to start from the simplest action in $d$ $=$ $4$ Minkowski
space which is  supersymmetric, super-gauge and $U(N)$ invariant. It can
written as
\beq
iS_{M}^{4} = \frac{i}{16}\int\! d^{4}xd^{2}\theta d^{2}\bar{\theta} \;\;
\bar{\Phi} \, e^V \Phi + \frac{i}{128g^{2}}\int\! d^{4}xd^{2}\theta
d^{2}\bar{\theta}\;\; VD\bar{D}^{2}DV  \label{4daction}
\eeq
where $\theta^{\alpha}$, $\bar{\theta}_{\dot{\alpha}}$ are complex spinors.
The spinor indices are raised and lowered through the totally antisymmetric
tensors $\epsilon^{\alpha\beta}$, $\epsilon^{\dot{\alpha}\dot{\beta}}$ (for a
review, see \cite{52}). The covariant derivatives are defined as
\begin{eqnarray}
D_{\alpha} &=& \frac{\partial}{\partial\theta^{\alpha}}
- i\sigma^{\mu}_{\alpha\dot{\alpha}}
\bar{\theta}^{\dot{\alpha}}\partial_{\mu}\;\;,
\hspace{1cm}\bar{D}^{\dot{\alpha}} =
\frac{\partial}{\partial\bar{\theta}_{\dot{\alpha}}}
+ i\theta ^{\alpha}\sigma^{\mu}_{\alpha\dot{\alpha}}\partial_{\mu} \nonumber \\
\sigma^{\mu}_{\alpha\dot{\alpha}} &=& (I,\vec{\sigma})\hspace{1cm}
\bar{\sigma}^{\mu,\dot{\alpha}\alpha} = (I,-\vec{\sigma})
\end{eqnarray}
where $\vec{\sigma}$ are the usual Pauli matrices which will induce the desired
two (three) dimensional euclidean $\gamma$ algebra. Recalling that $V$ is a
real superfield while $\Phi$ ($\bar{\Phi}$) is chiral
($\bar{D}_{\dot{\alpha}}\Phi$ $=$ $0$) and that one can fix a Wess-Zumino gauge
exactly in the same way as in our two dimensional model, we write (A.1) in
component form in order to see the field content and to perform the reduction.
We find it becomes
\bear
i\int d^{3}xdt\;\left[(\bar{D}_\mu z)(D^{\mu}z) + i\bar{\psi}\,
\bar{\Dslash}\
\psi + \bar{F}F \right] + \left[\lambda\bar{z}z + \bar{\psi}\bar{u}z
+ \psi u\bar{z}\right]
\enar
\beq
+ \, \frac{1}{f^{2}}\left[ -\frac{1}{4}F_{\mu\nu}F^{\mu\nu} + 2i\bar{u}
\bar{\partialslash}u
+ 2 d^{2}\right]_{\mbox{\footnotesize{irrelevant}}}
\eeq
where $D_{\mu}$ $=$ $\partial_{\mu}$ $-$ $iA_{\mu}$ and
$\bar{\Dslash}^{\dot{\alpha}\alpha}$ $=$
$D_{\mu}\bar{\sigma}^{\mu ,\dot{\alpha}\alpha}$.

The reason for the subscript $\mbox{irrelevant}$ will be clear in the
following. For the moment let us choose the euclidean time $it$ and the $x_{2}$
spatial direction as the dimensions to be reduced. A straightforward analysis
leads to the following two dimensional euclidean action:
\begin{eqnarray}
S_{E}^{(2)} &=& \int\!d^{2}_{E}x\;\left[(\bar{D}_i z)(D_{i}z)
+ i\bar{\psi}\bar{\Dslash}\psi - A_{0}^{2}\bar{z}z + A_{2}^{2}\bar{z}z
+ A_{0}\bar{\psi}\psi -  A_{2}\bar{\psi}\sigma_{2}\psi - \bar{F}F \right]
\\ \nonumber
&&~~~~ - \left[\lambda\bar{z}z + \bar{\psi}\bar{u}z + \psi u\bar{z}\right]
\\ \nonumber
&&~~~~ + \frac{1}{f^{2}}\left[ \frac{1}{4}F_{ij}F_{ij}
- \frac{1}{2}(\partial_{i}A_{0})(\partial _{i}A_{0})
+ \frac{1}{2}(\partial _{i}A_{2})(\partial _{i}A_{2})
- 2i\bar{u}\bar{\partialslash}u - 2 \lambda^{2}
\right]_{\mbox{\footnotesize{irrelevant}}}
\end{eqnarray}
It is clear that the first two pieces reproduces exactly the supersymmetric
$CP^{N}$ model with the components $A_{0}$ and $A_{2}$ which are nothing but
$\sigma$ and $\pi$ respectively while the two dimensional euclidean
$\gamma$-algebra is realized through the matrices ($\sigma_{1},\sigma_{3}$) and
$\gamma_{5}$ $=$ $\sigma_{2}$. The third term in square brackets, which was
generated by the kinetic term in (A.1), has dimension four in the $1/N$
expansion so that it is really an irrelevant operator in the approach to
criticality. Obviously we could have started with a (non-renormalizable) model
in $d$ $=$ $4$ perfectly analogous to $CP^{N}$ in its {$\NN$} $=$ $2$
formulation as in \cite{21}. However starting from a renormalizable model like
(A.1) gives the same insight into the $\gamma$-algebra structure and moreover
it highlights the similarity with the relation between $\phi^{4}$ and $O(N)$
$\sigma$ model in $2$ $\leq$ $d$ $<$ $4$, \cite{53}. The transition is obvious
and reducing only the temporal dimension we are left with in $3$ dimensions
\bear
S_{E}^{(3)} = \int\! d^{3}_{E}x\;\left[(\bar{D}_{i}z)(D_{i}z)
+ i\bar{\psi}\,\bar{\Dslash}\psi - A_{0}^{2}\bar{z}z + A_{0}\bar{\psi}\psi
- \bar{F}F \right] ~-~ \left[ \lambda\bar{z}z + \bar{\psi}\bar{u}z
+ \psi u\bar{z}\right]
\enar
\beq
+~ \frac{1}{f^{2}}\left[ \frac{1}{4}F_{ij}F_{ij}
- \frac{1}{2}(\partial _{\mu}A_{0})(\partial _{i}A_{0})
- 2i\bar{u}\bar{\partialslash}u
- 2 \lambda^{2}\right]_{\mbox{\footnotesize{irrelevant}}}
\eeq
The euclidean algebra is realized through the $\vec{\sigma}$ matrices and so no
$\pi$ is present as previously claimed.

\newpage
\epsfxsize= 17cm
\epsfbox{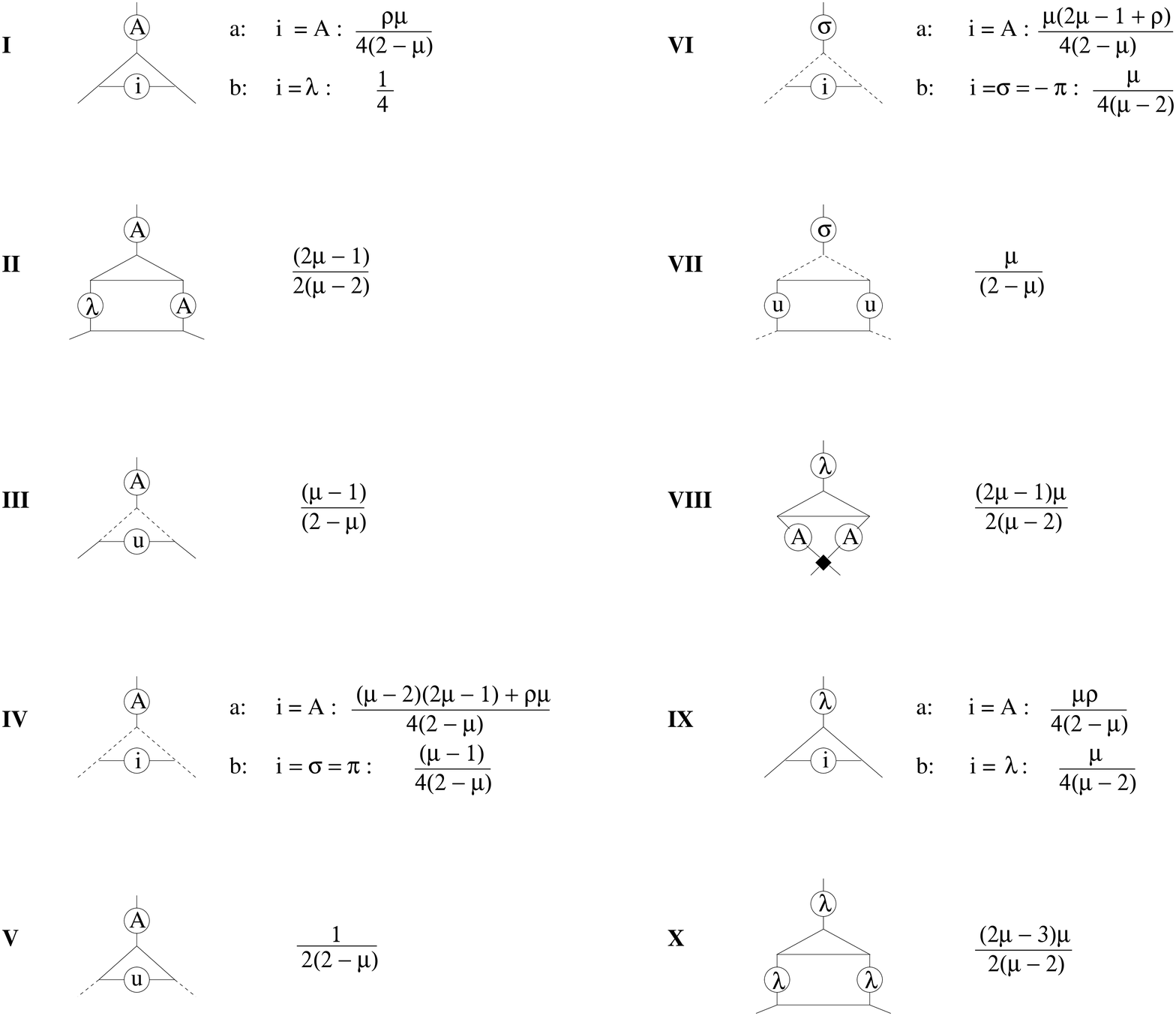}
\epsfxsize=17cm
\epsfbox{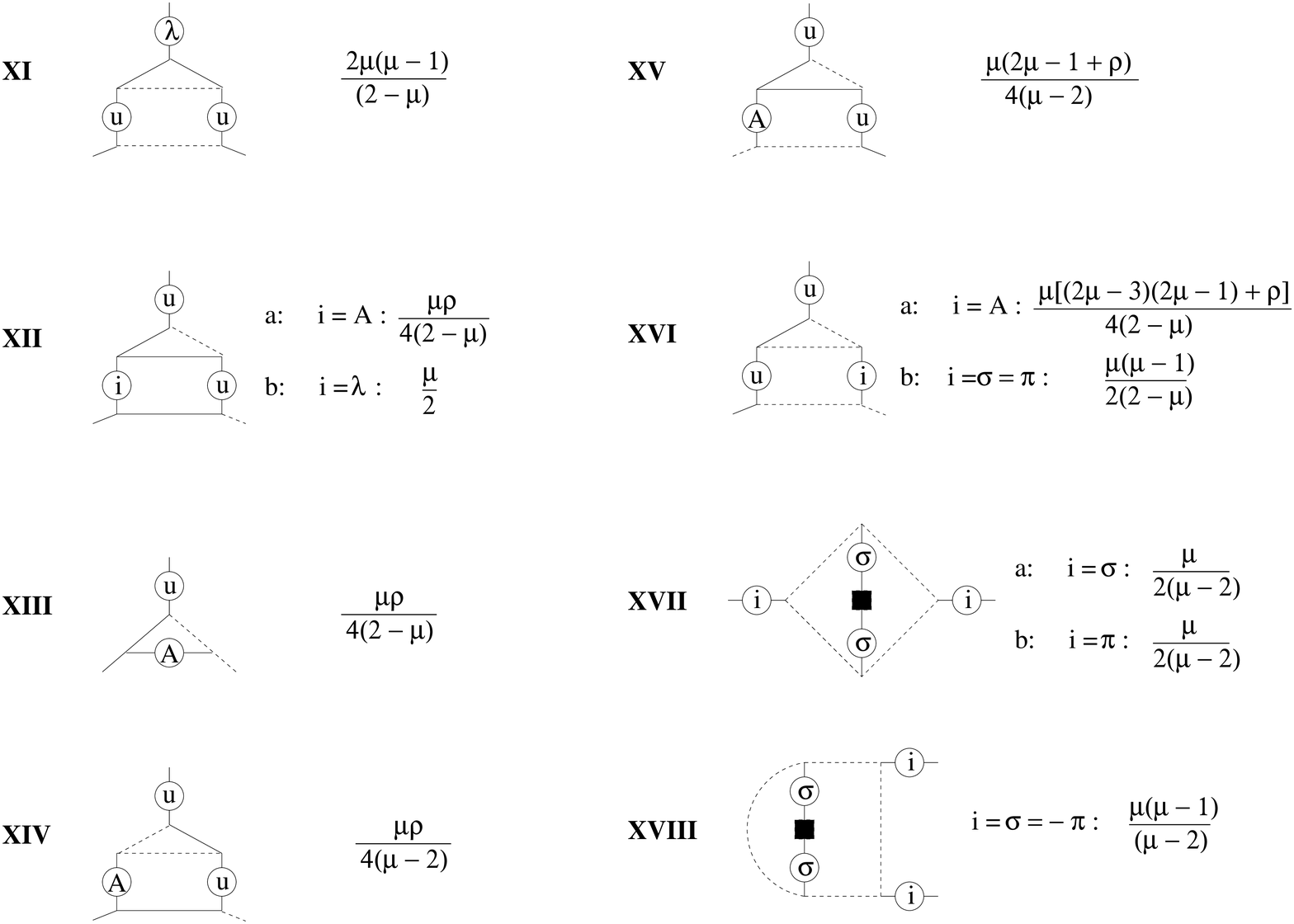}
\vspace{1cm}
\begin{center}
{\large {\bf Table 1. Pole parts for vertex graphs.}}
\end{center}
\newpage
\noindent
\epsfxsize = 15cm
\epsfbox{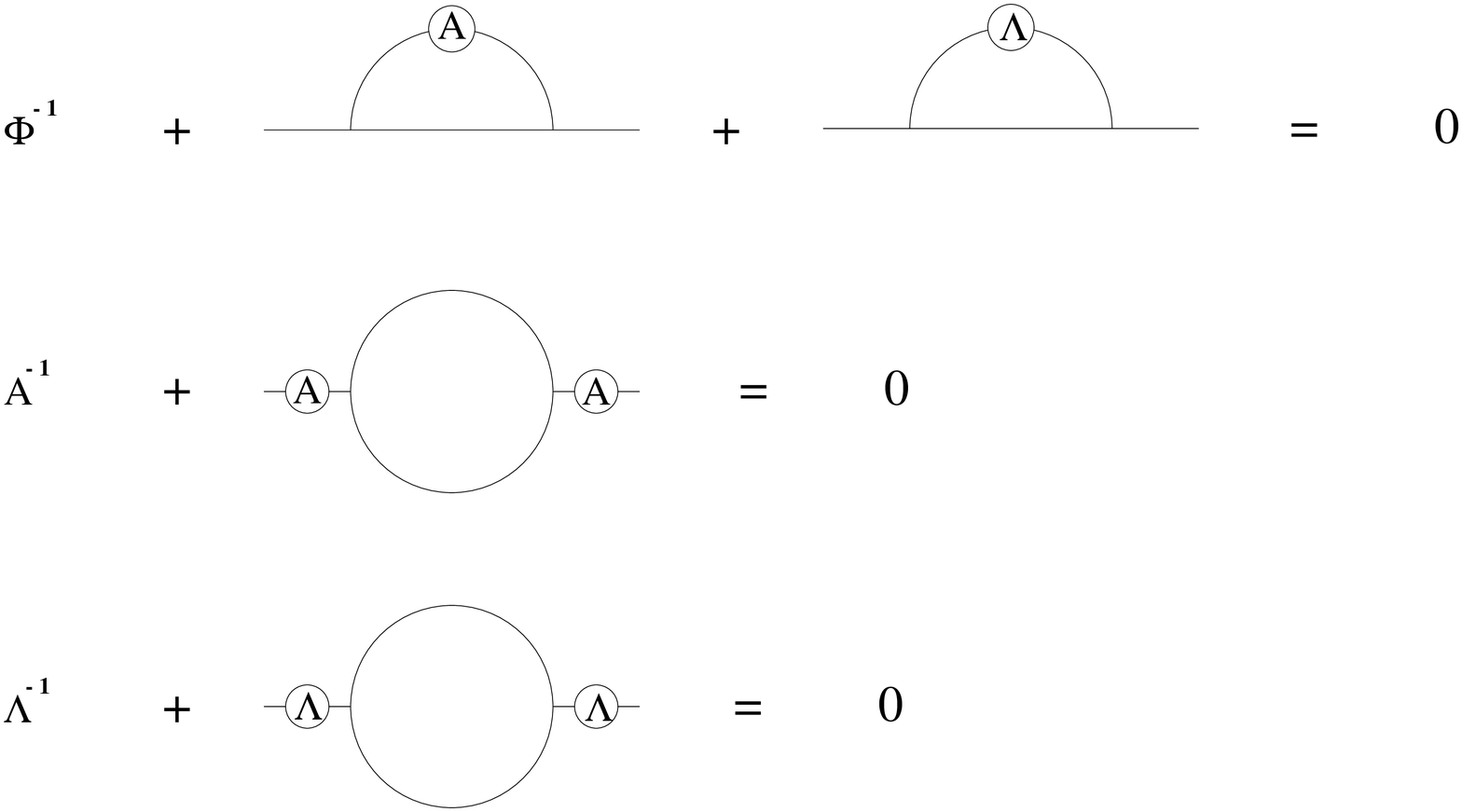}

\begin{center}
{\large {\bf Fig. 1. Dressed Schwinger Dyson equations for each field at
$O(1/N)$.}}
\end{center}
\vspace{4cm}

\epsfxsize= 12cm
\epsfbox{susycpnsign.eps}
\vspace{1cm}
\begin{center}
{\large {\bf Fig. 2. Extra graph contributing to $\sigma \bar{\psi} \psi$ in
$d$ $=$ $3$.}}
\end{center}
\newpage
\noindent
{\Large {\bf Figure Captions.}}
\begin{description}
\item[Fig. 1.] Dressed Schwinger Dyson equations for each field at $O(1/N)$.
\item[Fig. 2.] Extra graph contributing to $\sigma \bar{\psi} \psi$ in $d$ $=$
$3$.
\end{description}
\end{document}